\documentclass[sigconf,screen]{acmart}

\usepackage{enumitem}
\usepackage{mymacros}
\usepackage{lang}
\usepackage{tikz}
\usepackage{float}
\usepackage{subcaption}
\usepackage{multirow}
\usepackage{url}
\usepackage{pifont}
\usepackage{makecell}
\usepackage{mathpartir}
\newcommand{\SOLIDITYKEYWORDS}

\lstnewenvironment{solidity}
    {\lstset{%
      name=main
      language=Java,         
      columns=fullflexible,
      keepspaces=true,
      style=nonumbers,
      classoffset=0,
      frame=bt,
      deletekeywords={byte, char, int, float, double},
      keywordstyle=\color{blue}\bfseries,      
      classoffset=1,
      morekeywords={modifier, mapping, contract, require, uint, emit,
                    assert, if, else, contract, function, returns, var, public}, 
      keywordstyle=\color{brown}\bfseries,  
      classoffset=0,
      commentstyle=\color{blue},             
      morecomment=[l][\color{mauve}]{//},
      morecomment=[l][\color{red}]{//!},
      morecomment=[s][\color{mauve}]{/*}{*/},
      morecomment=[s][\color{red}]{/**}{*/},
      stringstyle=\color{dkgreen},             
      escapeinside={\%*}{*)}                   
}}{}

\lstnewenvironment{solidity-small}
    {\lstset{%
      name=main
      language=Java,         
      columns=fullflexible,
      keepspaces=true,
      style=tinynonumbers,
      classoffset=0,
      frame=bt,
      deletekeywords={byte, char, int, float, double},
      keywordstyle=\color{blue}\bfseries,      
      classoffset=1,
      morekeywords={modifier, mapping, contract, require, uint, emit,
                    assert, if, else, contract, function, returns, var, public}, 
      keywordstyle=\color{brown}\bfseries,  
      classoffset=0,
      commentstyle=\color{blue},             
      morecomment=[l][\color{mauve}]{//},
      morecomment=[l][\color{red}]{//!},
      morecomment=[s][\color{mauve}]{/*}{*/},
      morecomment=[s][\color{red}]{/**}{*/},
      stringstyle=\color{dkgreen},             
      escapeinside={\%*}{*)}                   
}}{}

\newcommand{\xmark}{\ding{55}}%

\newcommand{\ourtool}{\textsc{dyels}}%

\newcommand{\IndexType}[1]{\ensuremath{\mathit{#1}_{\mathit{I}}}}
\newcommand{\ConstructType}[1]{\ensuremath{\mathit{#1}}}

\newcommand{\VarStorIndex}[2]{#1 \ensuremath{\stackrel{S}{\to}} #2}
\newcommand{\VarValue}[2]{#1 \ensuremath{\to} #2}

\newcommand{\Flows}[2]{#1 \ensuremath{\rightsquigarrow} #2}

\newcommand{\StorageIndex}{\ensuremath{\mathit{SInd}}}
\newcommand{\ConstantIndex}[1]{\IndexType{Const}#1}
\newcommand{\MappingAccessIndex}[1]{\IndexType{Map}#1}
\newcommand{\ArrayDataStartIndex}[1]{\IndexType{ArrD}#1}
\newcommand{\ArrayAccessIndex}[1]{\IndexType{ArrA}#1}
\newcommand{\OffsetIndex}[1]{\IndexType{Offs}#1}

\newcommand{\ActualIndex}[1]{\ensuremath{\Downarrow} #1}

\newcommand{\PHI}[2]{#1 := \ensuremath{\phi}(#2)}
\newcommand{\PHITrCl}[2]{#1 := \ensuremath{\phi^{*}}(#2)}

\newcommand{\StorageConstruct}[1]{\ConstructType{SCons}#1}
\newcommand{\IsConstruct}[1]{\ensuremath{\downarrow} #1}

\newcommand{\Constant}[1]{\ConstructType{Const}#1}
\newcommand{\Mapping}[1]{\ConstructType{Map}#1}
\newcommand{\Array}[1]{\ConstructType{Arr}#1}
\newcommand{\Offset}[1]{\ConstructType{Offs}#1}
\newcommand{\Variable}[1]{\ConstructType{Var}#1}
\newcommand{\PackedVariable}[1]{\ConstructType{PVar}#1}

\newcommand{\VarHoldsBytes}[7]{\ensuremath{#2 \textbf{: } #1[#4:#5] \to #3[#6:#7]}}

\newcommand{\VarHoldsBytesSplitA}[5]{\ensuremath{#2 \textbf{: } #1[#4:#5] \to}}
\newcommand{\VarHoldsBytesSplitB}[3]{\ensuremath{#1[#2:#3]}}

\newcommand{\AnyStorageVar}[2]{#1: #2}

\newcommand{\MapToStorVar}[2]{#1 \ensuremath{\mmapsto} #2}

\newcommand{\StorageVarBytesUse}[5]{#2 : #3 := #1[#4 : #5]}

\newcommand{\StorageVarBytesDef}[6]{\ensuremath{#2, #3 : #1[#5 : #6] := #4}}

\newcommand{\AnyStorageVarBytes}[4]{#1: #2[#3 : #4]}

\newcommand{\mergefail}[1]{\ensuremath{\mathbin{\sqcup\!\!\!\diagup} #1}}

\newcommand{\AppendixA}{Appendix~\ref{sec:packed}}

\copyrightyear{2026}
\acmYear{2026}
\setcopyright{cc}
\setcctype{by}
\acmConference[ICSE '26]{2026 IEEE/ACM 48th International Conference on Software Engineering}{April 12--18, 2026}{Rio de Janeiro, Brazil}
\acmBooktitle{2026 IEEE/ACM 48th International Conference on Software Engineering (ICSE '26), April 12--18, 2026, Rio de Janeiro, Brazil}
\acmPrice{}
\acmDOI{10.1145/3744916.3773121}
\acmISBN{979-8-4007-2025-3/26/04}

\begin{document}

\title{Precise Static Identification of Ethereum Storage Variables}


\author{Sifis Lagouvardos}
\orcid{0000-0002-6233-1548}
\affiliation{%
  \institution{University of Athens}
  \city{Athens}
  \country{Greece}
}
\affiliation{%
  \institution{Dedaub}
  \city{Athens}
  \country{Greece}
}
\email{sifis.lag@di.uoa.gr}

\author{Yannis Bollanos}
\orcid{0009-0006-6905-9264}
\affiliation{%
  \institution{Dedaub}
  \city{Athens}
  \country{Greece}
}
\email{ybollanos@dedaub.com}

\author{Michael Debono}
\orcid{0009-0009-5900-7833}
\affiliation{%
  \institution{Friendly Maltese Citizens}
  \city{Msida}
  \country{Malta}
}
\email{mixy1@ctf.mt}

\author{Neville Grech}
\orcid{0000-0002-6790-2872}
\affiliation{%
  \institution{Dedaub}
  \city{Msida}
  \country{Malta}
}
\email{me@nevillegrech.com}

\author{Yannis Smaragdakis}
\orcid{0000-0002-0499-0182}
\affiliation{%
  \institution{University of Athens}
  \city{Athens}
  \country{Greece}
}
\affiliation{%
  \institution{Dedaub}
  \city{Athens}
  \country{Greece}
}
\email{smaragd@di.uoa.gr}


\begin{abstract}
  Smart contracts are small programs that run autonomously on the blockchain, using it as their
  persistent memory. The predominant platform for smart contracts is the Ethereum VM (EVM). In
  EVM smart contracts, a problem with significant applications is to identify data structures (in blockchain state, a.k.a. ``storage''),
  given only the deployed smart contract code. The problem has been highly challenging and has often been considered nearly
  impossible to address satisfactorily. (For reference, the latest state-of-the-art research tool fails to recover nearly
  all complex data structures and scales to 50\% of contracts.) Much
  of the complication is that the main on-chain data structures (mappings and arrays) have their
  locations derived dynamically through code execution.

  We propose sophisticated static analysis techniques to solve the
  identification of on-chain data structures with extremely high
  fidelity and completeness. Our analysis scales nearly universally and recovers deep data structures.
  Our techniques are able to identify the exact types of data structures
  with 95.70\% precision and at least 94.96\% recall, compared to a state-of-the-art
  tool managing 83.30\% and 55.65\% respectively.
  Strikingly, the analysis is often more complete than
  the storage description that the compiler itself produces, with full access to the source code.
\end{abstract}




\maketitle

\section{Introduction}
Smart contracts on programmable blockchains have been successfully used to implement complex applications,
mostly of a financial nature \cite{SoK:DeFi}. The dominant platform for smart contracts is the
Ethereum VM (EVM): the execution layer behind blockchains such as Ethereum, Arbitrum, Optimism, Binance, Base,
and many more. Millions of smart contracts have been deployed on these chains and can be invoked on-demand.
Many thousands of them are in active use every day.

To enable the persistence of data between different blockchain transactions, contracts employ the blockchain as
their persistent memory, to save their state. In EVM terms, this persistent memory is called
``\emph{storage}'' and is accessed using special random-access instructions. A challenge of high value
has emerged out of the use of storage in smart contracts: recovering high-level storage structures from
the deployed form of the smart contract, i.e., from EVM bytecode. This task is crucial for several applications:

\begin{itemize}
  \item \textbf{Security Analysis}: A number of smart contract vulnerabilities arise from incorrect handling of storage variables, such as storage collisions in upgradable contracts~\cite{NotYourType}. Precise modeling of storage is required for detecting such vulnerabilities.

  \item \textbf{Decompilation and Reverse Engineering}: Tools \cite{elipmoc,panoramix,heimdall} that decompile EVM bytecode back to high-level code rely on storage modeling to reconstruct variable declarations and data structures~\cite{VarLifter}.

  \item \textbf{Off-chain Applications}: Blockchain explorers, debuggers, and other off-chain tools need to interpret storage data to provide meaningful information to users. They often rely on compiler-generated metadata, which may be incomplete or unavailable~\cite{Etherscan} for interesting smart contracts like proprietary bots or hacker contracts.

  \item \textbf{Static Analysis and Verification}: Precise storage modeling enables advanced static analysis and formal verification of smart contracts, facilitating the detection of bugs and the proof of correctness~\cite{MadMax, ethainter,hyperion,10.1145/3660786}.
\end{itemize}

Smart contracts are written overwhelmingly
in the Solidity language,
which allows developers to define storage variables ranging from simple value types to complex, arbitrarily nested data structures such as arrays, mappings, and structs.

However, when Solidity code is compiled into EVM bytecode, much of the high-level structure and type information is lost. This is because the EVM's permanent storage is a simple key-value store mapping 256-bit keys to 256-bit values. Due to the luxury of having a large key space, the default pattern for high-level languages targeting the EVM is to translate high-level constructs into low-level storage access patterns using cryptographic hashing and arithmetic operations to compute storage slots dynamically. This transformation creates a significant gap between the high-level representation of storage variables and the low-level state of permanent storage reflected on the blockchain.

Existing approaches to storage modeling face significant limitations. Early frameworks~\cite{Albert2018,Tsankov:2018:SPS:3243734.3243780} often reasoned about storage operations only when storage indexes were constants, sacrificing precision or completeness when dealing with dynamic data structures. While some tools~\cite{MadMax, ethainter} introduced methods to infer high-level storage structures, they lacked support for arbitrarily nested data structures and complex storage patterns. Recent tools like VarLifter~\cite{VarLifter} attempt to recover storage layouts but struggle with scalability and completeness, failing to produce output for a substantial portion of real-world contracts.

\textbf{Contributions.} This paper introduces \ourtool{}, a static analysis approach that accurately infers high-level storage structures from EVM bytecode. Our key contributions are:

\begin{itemize}
  \item \textbf{Static Storage Modeling}: We develop a novel static analysis that fully supports arbitrarily nested composite data structures in Solidity.
  By employing a recursive storage analysis, \ourtool{} scalably and precisely reconstructs complex storage layouts from low-level bytecode.
  \item \textbf{Evaluation on All Deployed Contracts}: \ourtool{}'s scalable design allows us to evaluate its performance on enormous datasets.
  The first consists of 377,132 smart contracts with ground truth provided by the Solidity compiler and allows us to assess the faithfulness of \ourtool{}'s recovered inferences.
  The second includes 903,805 deduplicated contracts, corresponding to all 73 million non-empty contracts deployed on the Ethereum mainnet.\footnote{All numbers reported over deployed contracts are as of Jun. 28, 2025.} This enables scalable recognition of storage patterns on the whole blockchain.

  \item \textbf{Evaluation Against Existing Tools}: We find the current state-of-the-art tool, VarLifter~\cite{VarLifter}, to terminate on only 50.5\% of contracts.
    When successful, VarLifter misses over 40\% of storage variables, including most non-trivial structures. This performance underscores the difficulty of the problem being
    solved. In comparison, \ourtool{} analyzes 99.36\% of contracts and provides higher-fidelity results, recovering the vast majority
    (93.55\%) of storage structures with excellent precision (97.70\%).

  \item \textbf{Enhanced Completeness Beyond Compiler Metadata}: We show that \ourtool{} can infer storage variables and structures not present in compiler-generated metadata,
  particularly those involving low-level storage patterns common in upgradable contracts.
  Analyzing the entire blockchain, we were able to identify over 300,000 contract addresses making use of such widespread low-level variables.

\end{itemize}

The core of \ourtool{} is released as open-source software as part of the Gigahorse lifting toolchain
~\footnote{\url{https://github.com/nevillegrech/gigahorse-toolchain}}
and has seen significant adoption in both industry and academia.
Its output is integrated into the decompilation of \url{https://app.dedaub.com}, the Dedaub security
and decompilation tool suite used by over 10,000 registered users.

\section{Background}\label{sec:background}

We next provide background on the Ethereum Virtual Machine and its storage model.

\subsection{Ethereum and the Ethereum Virtual Machine}

Ethereum is a decentralized blockchain platform that enables the execution of smart contracts---autonomous
programs that run on the blockchain. Smart contracts are predominantly written in the Solidity high-level language and are compiled into bytecode for execution on the Ethereum Virtual Machine (EVM). The EVM has dominated as an execution platform and has been adopted by most other programmable blockchains.

The EVM is a stack-based virtual machine designed to execute smart contracts securely and deterministically.
It operates on 256-bit words, utilizes its own \emph{stack}, \emph{memory}, \emph{transient storage} and \emph{persistent storage},
and provides a Turing-complete execution environment. This paper focuses on \emph{persistent storage} or
just \emph{storage}.

\subsection{EVM Bytecode Format and Execution Model}

EVM bytecode is a sequence of instructions, each represented by a single-byte opcode (with an immediate argument for \sv{PUSH} opcodes).
The EVM supports a rich, albeit unconventional set of operations, which includes anything from arithmetic, logic, control flow, hashing, and state and memory interaction.

As a stack-based machine, most EVM opcodes perform computations using a stack of 256-bit words, when such an opcode has a fixed operand or return size.
The EVM also features different kinds of state. Memory is a \emph{dense} addressable byte array that is cleared at the end of each transaction with a smart contract (from the outside or nested, via calls from one contract to the next). A recent addition is transient storage, which is cleared at the end of the outermost transaction.


\subsection{Storage in the EVM}

At the EVM level, persistent storage (or simply \emph{storage}) is a persistent key-value store where both keys and values are 256-bit words.
Storage maintains the state of a contract between transactions. Storage is simply a sparse word array indexed by 256-bit words, spanning from keys $0$ to $2^{256}-1$.

High-level languages like Solidity provide structured data types such as integers, arrays, mappings, and structs.
The Solidity compiler maps these high-level constructs to EVM storage using specific patterns, now also widely-adopted in other compilers.

The EVM only allows reading from and writing to storage using the \sv{SLOAD} and \sv{SSTORE} instructions, both of which index storage using a 32-byte index value,
with the value read or written also being fixed at 32 bytes. As high-level languages need to implement arbitrarily complex data structures using such
very low-level primitives, there is a huge disparity between the source and bytecode representations.

\subsection{EVM Storage Model and Solidity Storage Layout}

During compilation, the Solidity compiler generally predictably orders smart contract storage variables using a deterministic heuristic
(e.g., by employing C3 linearization upon inheritance) and assigns a slot $p$ to each variable, and accumulates $p$ by an appropriate amount at each step.

The following cases briefly capture the mapping of high-level constructs to EVM storage:

\begin{itemize}

  \item \textbf{Value Types}: Simple value types (e.g., \sv{uint256}, \sv{bool}, \sv{address}) are stored in sequential storage slots starting from slot $0$.
  To optimize space, multiple small values may be packed into a single 32-byte storage slot. For example, two \sv{uint128} variables can share one slot,
  and smaller types like \sv{bool} and \sv{uint8} can be packed together.

  \item \textbf{Static Arrays}: Fixed-size arrays are stored by sequentially allocating storage slots for each element. For an array declared as \sv{T[n]}, where \sv{T}
  is the element type and \sv{n} is the fixed size, elements are stored starting from slot $p$, the assigned slot of the array variable.
  The element at index $i$ is stored at slot $p + i$.

  \item \textbf{Dynamic Arrays}: Dynamic arrays store their length at a fixed slot $p$, and their elements are stored starting from the Keccak-256 hash of slot $p$. (This is a cryptographic hash function, expected to be collision-resistant.)
  Specifically, element $i$ is stored at position $\mathit{keccak256}(p) + i$ where $p$ is the slot assigned to the array variable. This allows for arrays of
  arbitrary length without preallocating storage slots.

  \item \textbf{Mappings}: Mappings use a hashing scheme to avoid key collisions. A mapping declared as \sv{mapping(K => V)} at slot $p$ stores a value associated
  with key $k$ at $\mathit{keccak256}\left(\mathit{encode}(k) \parallel p\right)$ where $\parallel$ denotes concatenation, and $\mathit{encode}(k)$ is the padded representation of key $k$.
  This ensures that each key-value pair in the mapping has a unique storage slot.

  \item \textbf{Structs}: Structs are stored by sequentially allocating storage slots for each of their members, similar to value types.
  For a struct declared as \sv{struct S \{ ... \}} and a variable of type \sv{S} assigned to slot $p$, its members are stored starting from slot $p$.
  If a struct contains members that are arrays or mappings, the storage rules for arrays and mappings are applied recursively to those members.
\end{itemize}

The recursive nature of these storage rules makes static modeling of storage complex, especially when only the EVM bytecode is available.
For instance, mappings and dynamic arrays involve runtime computations of storage slots using hash functions, which are challenging to resolve statically.
Additionally, the packing of multiple variables into a single storage slot requires bit-level instruction analysis to accurately extract individual variables.


\section{Smart Contract Storage Patterns}\label{sec:compiler-patterns}

We next show a simple smart contract that will serve as an example to explain
how the most common data structures provided by Solidity are implemented in the
low-level EVM bytecode.

\begin{figure}[H]
  \begin{solidity}
contract StorageExample {
  uint256 public supply;  // slot 0x0
  address public owner;   // slot 0x1
  bool public isPaused;   // slot 0x1
  uint256[] public supplies; // slot 0x2
  mapping (address => bool) public admins; // slot 0x3
  struct vals {uint256 field0; uint256 field1;}
  mapping (address => mapping(uint256 => vals)) public complex; // slot 0x4
}\end{solidity}
  \caption{Example Smart Contract}
  \label{example}
\end{figure}

\subsection{Low-Level Implementation Patterns}
For our example contract in Figure~\ref{example}, the storage layout translates
to the following low-level operations, also shown schematically in Figure~\ref{fig:storage_layout}:

\begin{itemize}
  \item \textbf{Simple Values (slot 0x0)}: The \sv{supply} variable occupies a full slot:
        \begin{itemize}
          \item Load: \sv{SLOAD(0x0)}
          \item Store: \sv{SSTORE(0x0) = newSupply}
        \end{itemize}

  \item \textbf{Packed Values (slot 0x1)}: The \sv{owner} (20 bytes) and \sv{isPaused} (1 byte) share slot 0x1:
        \begin{itemize}
          \item Load owner: \sv{SLOAD(0x1)} followed by \sv{AND(loaded,}
                  \\\sv{0xffffffffffffffffffffffffffffffffffffffff)}
          \item Load \sv{isPaused}: \sv{SLOAD(0x1)} followed by \sv{AND(SHR(0xa0, loaded), 0xff)}.
          The masked variable is often followed by two \sv{ISZERO(ISZERO(masked))}.
          \item Store requires reading existing value, masking, and combining with new value.
        \end{itemize}

  \item \textbf{Dynamic Array (slot 0x2)}: For \sv{supplies}:
        \begin{itemize}
          \item Length access: \sv{length = SLOAD(0x2)}
          \item Element $i$ access: \sv{SLOAD}($\mathtt{keccak256}(\sv{0x2}) + i$)
          \item Store: \sv{SSTORE}($\mathtt{keccak256}(\sv{0x2}) + i$) \sv{=  value}
        \end{itemize}

  \item \textbf{Mapping (slot 0x3)}: For \sv{admins}:
        \begin{itemize}
          \item Key $k$ access: \sv{SLOAD}($\mathtt{keccak256}(\mathtt{pad32}(k) \parallel \sv{0x3})$)
          \item Store: \sv{SSTORE}($\mathtt{keccak256}(\mathtt{pad32}(k) \parallel \sv{0x3})$) \sv{ = value}
        \end{itemize}

  \item \textbf{Nested Mapping (slot 0x4)}: For \sv{complex}:
        \begin{itemize}
          \item For keys $k_1$, $k_2$:\\
          \sv{field0} is accessed using: \sv{SLOAD}($\mathtt{keccak256}(\mathtt{pad32}(k_2) \parallel \mathtt{keccak256}(\mathtt{pad32}(k_1) \parallel \sv{0x4}))$)\\
          \sv{field1} is accessed using: \sv{SLOAD}($\mathtt{keccak256}(\mathtt{pad32}(k_2) \parallel \mathtt{keccak256}(\mathtt{pad32}(k_1) \parallel \sv{0x4})) + 1$)
        \end{itemize}
\end{itemize}

\begin{figure}
  \centering
  \resizebox{\columnwidth}{!}{%
    \begin{tikzpicture}[thick,every node/.style={scale=1.25}]
      \draw[rounded corners=15pt] (0,0) rectangle (14,15);
      \node at (6,14.4) {Contract Storage};

      \draw[gray!30] (1,12.6) rectangle (4,13.4);
      \node[anchor=west] at (0,13) {1};

      \foreach \y/\l in {9.8/5,9/6,8.2/$\cdots$} {
        \node[anchor=west] at (0,\y) {\l};
        \draw[gray!30] (1,\y-0.4) rectangle (10,\y+0.4);
      }
      \begin{scope}
        \foreach \y/\l/\c in {
          13.8/0/{\sv{uint256}},
          12.2/2/{\sv{uint256[]}} ,
          11.4/3/{\sv{mapping(address => bool)}},
          10.6/4/{\sv{mapping(address => mapping(uint256 => uint256))}}
        } {
          \node[anchor=west] at (0,\y) {\l};
          \draw (1,\y-0.4) rectangle (10,\y+0.4);
          \node at (5.5,\y) {\c};
        }

        \node at (4.5, 13) {\sv{bool}};
        \draw (4,12.6) rectangle (5,13.4);
        
        \node at (8, 13) {\sv{address}};
        \draw (5,12.6) rectangle (10,13.4);


        \node[anchor=west] at (0,6.6) {$\cdots$};
        \draw[gray!30] (1,6.2) rectangle (10,7);

        \node[anchor=west] at (0,7.4) {\sv{0x17..d3ec}};
        \node at (5.5,7.4) {\sv{admins[0]}};
        \draw (1,7) rectangle (10,7.8);
        
        \node[anchor=west] at (0,4.2) {$\cdots$};
        \draw[gray!30] (1,3.8) rectangle (10,4.6);

        \node[anchor=west] at (0,5.8) {\sv{0xb0...d3b8}};
        \node at (5.5,5.8) {\sv{complex[0][0].field0}};
        \draw (1,5.4) rectangle (10,6.2);

        \node[anchor=west] at (0,5) {\sv{0xb0...d3b9}};
        \node at (5.5,5) {\sv{complex[0][0].field1}};
        \draw (1,4.6) rectangle (10,5.4);

        \node[anchor=west] at (0,1.8) {$\cdots$};
        \draw[gray!30] (1,1.4) rectangle (10,2.2);

        \node[anchor=west] at (0,3.4) {\sv{0xc2...f85b}};
        \node at (5.5,3.4) {\sv{supplies[0]}};
        \draw (1,3) rectangle (10,3.8);

        \node[anchor=west] at (0,2.6) {\sv{0xc2...f85c}};
        \node at (5.5,2.6) {\sv{supplies[1]}};
        \draw (1,2.2) rectangle (10,3);
        
        \node[anchor=west] at (0,1) {\sv{0xff...ffff}};
        \draw[gray!30] (1,0.6) rectangle (10,1.4);

        \draw[dashed] (10,12.2) -- (13.8,12.2) -- (13.8,3.4) -- (10,3.4)
          node[left, right, text width=2cm] {\sv{keccak256(slot) + index}};          
        
        \draw[dashed] (10,11.4) -- (13.35,11.4) -- (13.35,7.4) -- (10,7.4)
          node[left, right, text width=2.5cm] {\sv{keccak256(key $\parallel$ slot)}};          
        
        \draw[dashed] (10,10.6) -- (12.9,10.6) -- (12.9,5.8) -- (10,5.8)
          node[left, right, text width=3cm] {\sv{keccak256($k_2$ $\parallel$ keccak256($k_1$ $\parallel$ slot))}};    
      \end{scope}
    \end{tikzpicture}%
  }
  \caption{Low-level Storage Layout Implementation of our example in Figure~\ref{example}}
  \label{fig:storage_layout}
\end{figure}

\subsection{Low-level Storage Patterns in High-level Code}

Although high-level storage patterns allow developers to implement powerful protocols,
making use of complex high-level data structures, the storage allocation algorithm has its
drawbacks. Solidity does not offer a high-level way to override the assigned storage slot of a variable
declaring it at an arbitrary slot. This functionality is needed by various standards requiring
compatible storage layouts. The most important such standard is ERC-1967 \cite{eip-1967},
standardizing the allocated storage slots to be used for the implementation, admin,
and beacon contract addresses of widely-used~\cite{ProxyEx} upgradable proxy contracts.
To support these standard patterns developers make use of Solidity's inline assembly~\cite{InlineAssembly},
as shown in Figure~\ref{fig:lowLevel}.

\begin{figure}
\begin{solidity}
bytes32 internal constant _ADMIN_SLOT =
 0xb53127684a568b3173ae13b9f8a6016e243e63b6e8ee1178d6a717850b5d6103;

// returns the value of the address variable stored at _ADMIN_SLOT
function _getAdmin() internal view returns (address) {
  return StorageSlot.getAddressSlot(_ADMIN_SLOT).value;
}
struct AddressSlot { address value; }

// Returns an `AddressSlot` with member `value` located at `slot`.
function getAddressSlot(bytes32 slot) internal pure
 returns (AddressSlot storage r) {
  assembly {
    r.slot := slot
  }
}  
\end{solidity}
\caption{Low-level code implementing the ERC-1967 standard.}
\label{fig:lowLevel}
\end{figure}

Such low-level code patterns allow users to use storage variables that are not declared as such.
Thus, these variables are unknown to the Solidity compiler, and are not included in its storage layout metadata.
This incompleteness of the compiler-produced metadata will be examined in our evaluation of Section~\ref{sec:evaluation}.

\section{Analysis Preliminaries and Input}

The \ourtool{} approach is a static analysis of the program's (smart contract's) code
that identifies the low-level patterns that
the Solidity compiler produces to implement the high-level features presented in Section \ref{sec:compiler-patterns}.
The challenge is to maintain the right level of analysis precision and scalability/computability, since
the analysis needs to model the derivation structure of arbitrary dynamic numerical quantities.


\begin{figure}[]
  \begin{tabular}{l l}
    $V$: set of variables &
    $S$: set of program statements \\
    $C$: set of 256-bit numbers &
    \type{Int}: set of 16-bit numbers \\
  \end{tabular}
  \caption{Type domain definitions}
  \label{input-types}
  \end{figure}

\begin{figure}[]
  \begin{tabular}{|l|}
    \hline
    \mystmtid{I} \textbf{:} \statement{LOAD}{iv}{r} $|$ \myvar{r} $\in$ $V$, \mystmtid{I} $\in$ $S$, \myvar{iv} $\in$ $V$ \\
    \sv{SLOAD} statement \mystmtid{I} loads into \myvar{r} the value  of storage  location \\ pointed-to by \myvar{iv}  \\
    \hline
    \mystmtid{I} \textbf{:} \storestatement{STORE}{iv}{u} $|$ \mystmtid{I} $\in$ $S$, \myvar{iv} $\in$ $V$, \myvar{u} $\in$ $V$ \\
    \sv{SSTORE} statement \mystmtid{I} stores the value of \myvar{u}  into the storage \\ location pointed to by \myvar{iv}\\
    \hline
    \statement{ADD/SUB/MUL}{a, b}{r}  $|$ \myvar{r} $\in$ $V$, \myvar{a} $\in$ $V \cup C$, \myvar{b} $\in$ $V \cup C$ \\
    Binary arithmetic operation over variable or constant operands \\ \myvar{a, b} \\
    \hline
    \VarValue{v}{c} $|$ \myvar{v} $\in$ $V$, \myvar{c} $\in$ $C$ \\
    Constant folding and constant propagation analysis.\\ \myvar{v} has constant value \myvar{c}. \\
    \hline
    \PHI{\myvar{l}}{\myvar{u}} $|$ \myvar{l} $\in$ $V$, \myvar{u} $\in$ $V$ \\
    SSA PHI instructions \\
    \hline
    \statement{HASH}{\myvar{a}, *}{r} $|$ \myvar{r} $\in$ $V$, \myvar{a} $\in$ $V$ \\
    \sv{SHA3} operation that computes the keccak256 hash \\of a variable number of args, storing it into \myvar{r} \\
    \hline
  \end{tabular}
  \caption{Input relation definitions}
  \label{rels}
  \end{figure}

Figures \ref{input-types} and \ref{rels} define the input schema for our analysis. (In using these predicates, we drop elements
that are not needed for the rule at hand, e.g., we may write ``\statement{LOAD}{iv}{r}'' instead of ``\textit{I} \textbf{:} \statement{LOAD}{iv}{r}'' when the
instruction identifier \textit{I} is unused.)

While these types and relations originate from the Elipmoc/Gigahorse lifter toolchain \cite{gigahorse,elipmoc,shrnkr,ethereum-memory-oopsla}, our approach is not restricted to this framework.
Instead, it can be applied to any mature decompilation framework that lifts the stack-based EVM bytecode into a register-based static-single-assignment (SSA) representation.

Some relations in Figure~\ref{rels} directly correspond to the register-based representation, such as \sv{LOAD} and \sv{STORE},
while others, like \sv{ADD}, \sv{SUB}, and \sv{MUL}, also incorporate the results of a constant folding and constant propagation analysis.

Finally, the \sv{HASH} relation, which supports the low-level implementation patterns presented in Section~\ref{sec:compiler-patterns}, stems from the EVM ``memory'' analysis~\cite{ethereum-memory-oopsla} built on top of Gigahorse.
While this specific implementation is tied to Gigahorse, similar EVM memory analyses~\cite{CertoraMemory,AlbertMemory} have been developed on top of other SSA-based analysis frameworks.

\section{Structure Identification}\label{sec:structure-identification}

The \ourtool{} analysis has two main parts: a) discovering the \emph{structure} of a program's storage layout (e.g.,
which structures are nested arrays or mappings); b) discovering the \emph{types} of data stored in every entry of each
structure. This section presents the first part: how to identify the data structures in a smart
contract's storage.

\subsection{Storage Index Value-flow Analysis}

The backbone of the analysis is a value-flow analysis that computes the values of all potential storage
index expressions,
and then uses the ones that end up being used in \emph{actual} storage operations to identify the constructs in the
program's storage layout.

\begin{figure}[]
  \begin{tabular}{lcll}
  type \StorageIndex{} &=&\ConstantIndex{}&(c: $C$)\\
  & $|$ & \ArrayAccessIndex{}&(par: \StorageIndex, iv: $V$)\\
  & $|$ & \ArrayDataStartIndex{}&(par: \StorageIndex)\\
  & $|$ & \MappingAccessIndex{}&(par: \StorageIndex, kv: $V$)\\
  & $|$ & \OffsetIndex{}&(par: \StorageIndex, of: \type{Int})
 \end{tabular}
  \caption{Vocabulary of storage index value expressions}
  \label{index-type}
  \end{figure}

Figure~\ref{index-type} presents our definition of the storage index values for our value-flow analysis.
The Algebraic Data Type (ADT) in the figure aims to capture accesses to Solidity's arbitrarily-nested high-level structures.
The ADT effectively defines what the analysis can infer about potential storage indexes.

\ConstantIndex{} is the only non-recursive kind of \StorageIndex{} type. Every storage
index will include a \ConstantIndex{} as the leaf of its ADT value, since 
all high-level storage structures are assigned a constant offset by the compiler.
The rest of the storage index types are recursively built on top of a pre-existing \StorageIndex{} instantiation, encoded as \type{par} (``par'' for ``parent'').
These include \ArrayAccessIndex{} and \ArrayDataStartIndex{} used to model operations on dynamic arrays, \MappingAccessIndex{} for mapping operations,
and \OffsetIndex{}, which enables supporting struct accesses. In addition to the \type{par} index, \StorageIndex{} values that model an index to a
high-level data structure (array or mapping) also include the index or key variables.

To compute the possible storage indexes for arbitrarily-nested data structures we define our analysis as a set of recursive inference rules.
The rules are faithfully transcribed from a fully mechanized implementation, so they should be precise, modulo mathematical shorthands used
for conciseness.

The analysis first computes two new relations.

\begin{center}
\begin{tabular}{|l|}
\hline
\VarStorIndex{\myvar{v}}{\myvar{si}} $\big|$ \myvar{v} $\in$ V,  \myvar{si} $\in$ \StorageIndex \\
\textbf{Storage Index Overapproximation}: \\Variable \myvar{v} holds potential storage index \myvar{si}. \\
\hline
\ActualIndex{\myvar{si}} $|$   \myvar{si} $\in$ \StorageIndex \\
\textbf{Actual Storage Index}: \\\myvar{si} ends up being used in a storage loading/storing operation.\\
\hline
\end{tabular}
\end{center}

\subsubsection{Storage Index Overapproximation}
Figure \ref{fig:indexes} contains our analysis logic for overapproximating the possible storage index values. 

\begin{figure}
  \[
  \textsc{(Base)}
  \quad
  \inferrule{
      \mathsf{v} \to \mathsf{c}
  }{
      \VarStorIndex{\mathsf{v}}{\ConstantIndex{(\mathsf{c})}}
  }
  \]
  
  \[
  \textsc{(Mapping)}
  \quad
  \inferrule{
      \VarStorIndex{\mathsf{pv}}{\mathsf{si}} \\
      \mathsf{v} := \mathsf{HASH}(\mathsf{kv}, \mathsf{pv})
  }{
      \VarStorIndex{\mathsf{v}}{\MappingAccessIndex{(\mathsf{si}, \mathsf{kv})}}
  }
  \]
  
  \[
  \textsc{(Array Data)}
  \quad
  \inferrule{
      \VarStorIndex{\mathsf{pv}}{\mathsf{si}} \\
      \mathsf{v} := \mathsf{HASH}(\mathsf{pv})
  }{
      \VarStorIndex{\mathsf{v}}{\ArrayDataStartIndex{(\mathsf{si})}}
  }
  \]
  
  \[
  \textsc{(Array Access)}
  \quad
  \inferrule{
      \VarStorIndex{\mathsf{pv}}{\ArrayDataStartIndex{(\mathsf{si})}} \quad
      \mathsf{v} := \mathsf{ADD}(\mathsf{pv}, \mathsf{i}) \\
      \mathsf{i} := \mathsf{MUL}(\mathsf{iv}, \mathsf{c}) \\
      \mathsf{iv} : V \quad \mathsf{c} : C
  }{
      \VarStorIndex{\mathsf{v}}{\ArrayAccessIndex{(\mathsf{si}, \mathsf{iv})}}
  }
  \]
  
  \[
  \textsc{(Offs1)}
  \inferrule{
      \VarStorIndex{\mathsf{pv}}{\mathsf{si}} \quad
      \mathsf{v} := \mathsf{ADD}(\mathsf{pv}, \mathsf{c}) \quad
      \mathsf{si} : \ArrayAccessIndex{} \mid \MappingAccessIndex{} \quad
      \mathsf{c} : \type{Int}
  }{
      \VarStorIndex{\mathsf{v}}{\OffsetIndex{(\mathsf{si}, \mathsf{c})}}
  }
  \]
  
  \[
  \textsc{(Offs2)}
  \inferrule{
      \VarStorIndex{\mathsf{pv}}{\OffsetIndex{(\mathsf{si}, \mathsf{o})}} \\
      \mathsf{v} := \mathsf{ADD}(\mathsf{pv}, \mathsf{c}) \quad
      \mathsf{c} : \type{Int}
  }{
      \VarStorIndex{\mathsf{v}}{\OffsetIndex{(\mathsf{si}, \mathsf{c} + \mathsf{o})}}
  }
  \]
  \caption{Inference Rules for Storage indexes}
  \label{fig:indexes}
  \end{figure}

We start with the simpler cases of the analysis for inferring the structure of storage indexes, with
detailed explanation, to also serve as introduction to the meaning of inference rules and of the input schema.

The \textsc{``Base''} rule produces the initial set of possible storage indexes by considering the facts of the constant folding and constant propagation
analysis provided by the Gigahorse/Elipmoc framework. Per the input schema, ``\VarValue{}{}'' is the predicate capturing the
result of the constant propagation/folding, matching an IR variable (if it always holds a constant value) to its
value. This means that every static constant in the contract code will be considered as a possible constant storage index, to be used either as-is or as a building
block of more complex indexes.



The \textsc{``Mapping''} rule models mapping accesses with the help of the \sv{HASH} predicate provided by the EVM ``memory'' modeling
analysis.
The rule states that:
\begin{itemize}
\item if a variable, $\mathsf{pv}$ points to a likely storage index $\mathsf{si}$,
\item and if its concatenation to the contents of another variable, $\mathsf{kv}$, is hashed
  in the smart contract code (using the EVM's hash operation),
\item then the hash result variable will hold a \MappingAccessIndex{} (mapping access index) over $\mathsf{si}$ and $\mathsf{kv}$ is
  the key variable of the modeled mapping access operation.
\end{itemize}

The next two rules model dynamic arrays in storage. The storage locations of a dynamic
array are determined by first hashing an array identifier, and then performing index arithmetic
via addition and multiplication.



Similarly to the case of mappings, the \textsc{``Array Data''} rule will create a new index value pointing to the start of an array after inferring a \sv{HASH}
operation that hashes the contents
of a variable, and that variable points to a pre-existing index.

The second rule (\textsc{``Array Access''}) will infer that if a variable holding an \ArrayDataStartIndex{} is
added to the result of the multiplication of a variable and a constant, the variable defined by the addition operation will point to a new
\ArrayAccessIndex{} value, inheriting the parent index of the \ArrayDataStartIndex{} value and using the multiplied variable as its access/indexing value.

It is worth asking whether the above code patterns are \emph{always} indicating a dynamic array, or
could arise for random code. If the compiled code has been produced via compilation, these patterns
are very unlikely to arise for non-array structures. There is no other data structure with both
contiguous (indicated via addition) and regular (indicated via multiplication with a constant)
storage location access. Furthermore, the presence of a hashed value, via a 1-argument hash
operation, adds even more confidence to the inference. Finally, the potential over-approximation of storage
indexes will be, in the next step of the analysis, checked against the use of the index as a proper
array index. All these elements contribute to a very high-fidelity inference.

Finally, the last two rules create \OffsetIndex{} values that are used to model struct accesses in EVM storage.
A struct is being accessed by addition of constant field offsets to a base storage index.
The base storage index is that of a mapping or array.
(If it is a mere constant index, then there is no way to distinguish the struct from just an explicit listing of its fields.)

The first rule will create a new \OffsetIndex{} when a small integer is added to a variable pointing to a \MappingAccessIndex{}
or \ArrayAccessIndex{} value, while the second one recursively creates new \OffsetIndex{} values for additions of existing \type{OffsetIndex}
values and small integers.



\subsubsection{Filtering Out Non-Realized Indexes}
\label{sec:filtering}

The next step for computing a smart contract's storage layout is to identify the subset of indexes computed in the overapproximating \smash{\VarStorIndex{}{}} relation that
are \textit{actually} used in storage operations. Relation \ActualIndex{} is used to compute these storage indexes as shown in the rules of Figure~\ref{fig:actual-used}.




\begin{figure}


  \[
  \textsc{(A1)}
  \inferrule{
      \VarStorIndex{\mathsf{v}}{\mathsf{si}} \quad
      \big(\_ := \mathsf{LOAD}(\mathsf{v}) \lor \mathsf{STORE}(\mathsf{v}) := \_ \big)
  }{
      \ActualIndex{\mathsf{si}}
  }
  \]
  
  \[
  \textsc{(A2)}
  \inferrule{
      \VarStorIndex{\mathsf{sv}}{\mathsf{si}} \quad
      \PHITrCl{\mathsf{lv}}{\mathsf{sv}} \quad
      \big(\_ := \mathsf{LOAD}(\mathsf{lv}) \lor \mathsf{STORE}(\mathsf{lv}) := \_\big)
  }{
      \ActualIndex{\mathsf{si}}
  }
  \]
  
  \[
  \textsc{(A3)}
  \inferrule{
      \ActualIndex{\mathsf{si}} \quad \\
      \mathsf{si} : \ArrayAccessIndex{} \mid \ArrayDataStartIndex{} \mid 
      \MappingAccessIndex{} \mid \OffsetIndex{}
  }{
      \ActualIndex{\mathsf{si.par}}
  }
  \]
  \caption{Inference Rules for recognizing actual (used) storage indexes}
  \label{fig:actual-used}
  \end{figure}


The first two rules are inferring the end-level storage indexes when they are used in \sv{LOAD/STORE} operations either directly
or through PHI operations via the $\phi^*$ relation, the transitive closure of the $\phi$ relation of our input.
The last rule is introduced to transitively infer that all parent indexes of
``actual'' storage indexes are considered ``actual'' indexes as well.

This seemingly very simple logic hides an important subtlety. This concerns the treatment of
PHI ($\phi$) instructions: the data-flow merge instructions in a static-single-assignment (SSA)
representation.
PHI instructions are merging the values for the same higher-level variable that arrive,
via different program paths, to a control-flow merge point.
For instance, if a high-level program variable \sv{x} is set in two different branches (``then'' or ``else'')
of an \sv{if} statement, then \sv{x} is produced by a PHI whose arguments are the \sv{x}-versions in the
branches that merge.

Note that, in the rules we have seen (Figure~\ref{fig:indexes}), the left-hand-side variable of a
PHI instruction does \emph{not} become the first part of a \VarStorIndex{}{} entry, even if
the right-hand-side variables (one or multiple) \emph{are} in it.
Doing so would result in analysis non-termination. A PHI may be merging different potential indexes,
all captured at run-time by the same variable. Then if the variable cyclically feeds into itself (as in
the case of code with a loop), we would end up with an unbounded number of potential storage index
inferences.

This is the importance of the ``\textsc{A2}'' rule, handling PHI instructions.
Although PHI instructions do not yield more storage indexes, recognized storage indexing patterns
propagate through the transitive closure of PHI instructions. In this way, we can get the confidence
of recognizing actual storage indexes, without attempting to fully track them at every point in the
program.


\subsubsection{Storage index analysis results on our example}

Now that we have presented how the storage indexes are computed it is interesting to see the results of the actual index (\ActualIndex{})
relation for our example in Figure~\ref{example}:

\begin{datalogcode}
ConstI(0x1)
ConstI(0x0)
ConstI(0x2)
ConstI(0x3)
ConstI(0x4)
ArrayAI(ConstI(0x2), 0x14d)
MapI(ConstI(0x3), 0x11e)
MapI(ConstI(0x4), 0x80)
MapI(MapI(ConstI(0x4), 0x80), 0x8e)
OffsI(MapI(MapI(ConstI(0x4), 0x80), 0x8e), 1)
\end{datalogcode}

As can be seen, the computed actual storage indexes are constant indexes \sv{0x0} (containing the 256-bit \sv{supply} variable),
\sv{0x1} (containing variables \sv{owner} and \sv{isPaused}), and composite indexes to access array \sv{supplies} at index \sv{0x2}
and mappings admins and complex at indexes \sv{0x3} and \sv{0x4}, along with their parent indexes.


\subsection{Inferring Storage Constructs from Storage Index Values}

Figure~\ref{construct-type} presents our definition of the \StorageConstruct{} (storage construct) Algebraic Data Type, used to describe all data structures that can be found in
Solidity smart contracts. The constructor cases of \StorageConstruct{} cover the different \StorageIndex{} types, while also introducing \Variable{} as an option.
Instances of \Variable{} express a value-typed fundamental unit of data at the end of our nesting chain.
This can either be a top-level value-typed variable, the element of an array, the key to a mapping, or a struct member.
Finally, the \PackedVariable{} type is used to express a ``packed'' variable: a construct that takes up part of a 32-byte storage word.

\begin{figure}[]
  \begin{tabular}{lcll}
  type \StorageConstruct{} &=&\Constant{}&(c: $C$)\\
  & $|$ & \Array{}&(par: \StorageConstruct{})\\
  & $|$ & \Mapping{}&(par: \StorageConstruct{})\\
  & $|$ & \Offset{}&(par: \StorageConstruct{}, of: \type{Int})\\
  & $|$ & \Variable{}&(par: \StorageConstruct{})\\
  & $|$ & \PackedVariable{}&(par: \StorageConstruct{}, b: (\type{Int}, \type{Int}))
  \end{tabular}
  \caption{Storage Construct Type}
  \label{construct-type}
\end{figure}

To define the algorithms that identify the program's high-level structures we need to introduce the following additional notation/computed predicates.

\begin{center}
  \begin{tabular}{|l|}
  \hline
  \StorageConstruct{($\mathsf{si}$)} $|$ $\mathsf{si}$ $\in$ \StorageIndex{} \\
  \StorageConstruct{} constructor, syntactically translating\\ corresponding \StorageIndex{} cases. \\ 
  \hline
  \IsConstruct{$\mathsf{sc}$}  $|$ $\mathsf{sc}$ $\in$ \StorageConstruct{}{} \\
  Relation containing all storage constructs in a program.\\
  \hline
  \MapToStorVar{\textit{I}}{$\mathsf{sv}$}  $|$ \textit{I} $\in$ $S$, $\mathsf{sv}$ $\in$ \Variable{}  \\
  Relation mapping storage \sv{LOAD/STORE} instructions to\\ the storage variable they operate on.\\
  \hline
  \end{tabular}
\end{center}

Following the computation of the \ActualIndex{} ``used storage index'' relation we can populate the \IsConstruct{}
relation with all program structures, as shown in Figure \ref{fig:IsConstruct}.



\begin{figure}[hb]
\[
\textsc{(Base)}
\inferrule{
    \ActualIndex{}{\mathsf{si}}
}{
    \IsConstruct{\StorageConstruct{(\mathsf{si})}}
}
\]

\[
\textsc{(Var)}
\inferrule{
    \ActualIndex{}{\mathsf{si}} \quad
    \nexists \mathsf{si'} : (\ActualIndex{}{\mathsf{si'}} \quad \\
    \StorageConstruct{(\mathsf{si'})} = Arr(\StorageConstruct{(\mathsf{si})})  \lor
    \StorageConstruct{(\mathsf{si'})} = Map(\StorageConstruct{(\mathsf{si})})
    )
}{
    \IsConstruct{\textit{Var(}\StorageConstruct{(\mathsf{si}))}}
}
\]

\caption{Using the ``used storage index'' inferences to compute a program's storage constructs.}
\label{fig:IsConstruct}
\end{figure}

The first rule considers all constructs that were translated from storage indexes.
The second one introduces new \Variable{} instances for every translated construct that is
never used as a parent index to a more complex construct.

Finally, in Figure \ref{fig:stmtMap}, we map the \sv{LOAD} and \sv{STORE} statements to the instance of \Variable{}
they operate on.

\begin{figure}
  \[
  \inferrule{
    \IsConstruct{\textit{Var(}\StorageConstruct{(\mathsf{si}))}} \quad
    \VarStorIndex{\mathsf{v}}{\mathsf{si}} \\
    \big( \mathit{I}: \_ := \mathsf{LOAD}(\mathsf{v}) \quad \lor \quad \mathit{I}: \mathsf{STORE}(\mathsf{v}) := \_ \big)
  }{
      \MapToStorVar{\mathit{I}}{\textit{Var(}\StorageConstruct{(\mathsf{si}))}}
  }
  \]
  \[
  \inferrule{
    \IsConstruct{\textit{Var(}\StorageConstruct{(\mathsf{si}))}} \quad
    \VarStorIndex{\mathsf{v}}{\mathsf{si}} \quad
    \PHITrCl{\mathsf{u}}{\mathsf{v}} \\
    \big( \mathit{I}: \_ := \mathsf{LOAD}(\mathsf{u}) \quad \lor \quad \mathit{I}: \mathsf{STORE}(\mathsf{u}) := \_ \big)
  }{
      \MapToStorVar{\mathit{I}}{\textit{Var(}\StorageConstruct{(\mathsf{si}))}}
  }
  \]
  \caption{Mapping \sv{LOAD} and \sv{STORE} statements to the storage variables they operate on.}
  \label{fig:stmtMap}
\end{figure}

\subsubsection{(Packed) Variable Partitioning}
After computing the contract's storage construct instances we need to identify
instances of multiple variables packed together into the same 32-byte storage word.
These instances of \Variable{} are encoded as \PackedVariable{} (``packed variable'').

This analysis step, presented in \AppendixA, models all reads and writes
of each \Variable{} instance through low-level bit-masking and shifting operations.
If all read and write operations write to non-conflicting sub-word segments, \PackedVariable{}
instances are introduced, replacing the pre-existing \Variable{} inferences.

\subsubsection{Storage construct \Variable{} and \PackedVariable{} inferences on our example}
At this point in our analysis pipeline the following \Variable{} inferences will be produced
for our example in Figure~\ref{example}, each corresponding to a (potentially packed) top-level variable,
array element, mapping value, or struct member:

\begin{datalogcode}
Var(Const(0x0))           // uint256 supply
PVar(Const(0x1), 0, 19)   // address owner
PVar(Const(0x1), 20, 20)  // bool isPaused
Var(Array(Const(0x2)))    // uint256[] supplies
Var(Map(Const(0x3)))      // mapping admins
// the 2 fields of struct value of nested mapping complex
Var(Map(Map(Const(0x4)))) 
Var(Offs(Map(Map(Const(0x4))), 1)) 
\end{datalogcode}

\section{Value Type Inference}

The second part of the \ourtool{} analysis is to identify the types of data structure entries, i.e.,
the types of the \Variable{} (and \PackedVariable{}) instances identified in the structure recognition of the
previous section.




Once \Variable{} and \PackedVariable{} instances have been identified, type inference over them is primarily an
instance of inferring monomorphic types by process of elimination, based on compatible operations.

The value-types supported by Solidity are the following:
\begin{itemize}
  \item \sv{uintX} with \sv{X in range(8, 256, 8)} (all numbers from 8 to 256, for each increment of 8): Unsigned integers, left-padded
  \item \sv{intX} with \sv{X in range(8, 256, 8)}: Signed integers, left-padded
  \item \sv{address}: Address type, 20 bytes in width, left-padded
  \item \sv{bool}: Boolean, left padded
  \item \sv{bytesX} with \sv{X in range(1, 32, 1)}: Fixed width bytearrays, right-padded
\end{itemize}

\begin{table}
  \caption{Kinds of operations supported by each value type with the corresponding EVM instructions implementing them}
  \label{tab:types}
\begin{tabular}{|l|ccccc|}
\toprule
  ops &            \sv{bytesX}                   &                                \sv{uintX} &                                   \sv{intX} & \sv{address} & \sv{bool} \\
\midrule
   \small{equal}  &                \sv{EQ, SUB}              &                              \sv{EQ, SUB} &                                \sv{EQ, SUB} & \makecell{\sv{EQ,}\\ \sv{SUB}} & \makecell{\sv{EQ,}\\ \sv{SUB}} \\
\hline
 \small{logical}     &               \xmark                     &                            \xmark         &                               \xmark        &  \xmark      & \sv{ISZERO} \\
\hline
 \small{comp} &            \sv{LT, GT}                   &                               \sv{LT, GT} &                               \sv{SLT, SGT} &  \makecell{\sv{LT,}\\ \sv{GT}} & \xmark \\
\hline
   \small{bitwise}   & \makecell{\sv{AND, OR,}\\ \sv{XOR, NOT}} &   \makecell{\sv{AND, OR,}\\ \sv{XOR, NOT}} & \makecell{\sv{AND, OR,}\\ \sv{XOR, NOT}} &  \xmark & \xmark \\
\hline
     \small{shifts}  & \makecell{\sv{SHL, SHR,}\\ \sv{MUL, DIV}} & \makecell{\sv{SHL, SHR,}\\\sv{MUL, DIV}} & \makecell{\sv{SHL, SAR}} &  \xmark & \xmark \\
\hline
\small{arithm}   &                      \xmark              & \makecell{\sv{ADD, SUB,}\\\sv{MUL, DIV,} \\ \sv{MOD, EXP,}\\ \sv{ADDMOD} \\ \sv{MULMOD}} & \makecell{\sv{ADD, SUB,}\\ \sv{MUL, EXP,}\\ \sv{SMOD,} \\ \sv{SDIV}} &   \xmark & \xmark \\
\hline
\small{byte ind} &              \sv{BYTE}                  &                    \xmark            &                         \xmark         &  \xmark & \xmark \\
\bottomrule
\end{tabular}
\end{table}

 Table~\ref{tab:types} captures the \ourtool{} systematic encoding of the different high-level operations Solidity supports for its value types,
along with the low-level EVM instructions that implement them. In addition to the table, \sv{bool} typed variables
support the high-level short-circuiting \sv{\&\&} and \sv{||} operators, supported via control-flow patterns (i.e.,
with no single corresponding low-level EVM instruction).

In most cases, a simple analysis can identify a storage variable's type, given that the
packed variable partitioning analysis of the previous subsection will give us its width.
If a tightly packed variable is then moved to the leftmost bytes of a variable (i.e., is right-padded)
we identify its type as \sv{bytesX}.

For packed variables that are moved to the stack as left-padded variables, we can easily distinguish
signed- and unsigned-integer-typed variables as the former will be used in signed arithmetic operations,
after getting the variable's length extended to 256 bits via the \sv{SIGNEXTEND} operation.

The cases that remain ambiguous require further analysis to correctly infer the variable type. These
cases of ambiguity are:
\begin{itemize}
  \item \sv{bool} vs. \sv{uint8}
  \item \sv{address} vs.  \sv{uint160}
  \item \sv{uint256} vs. \sv{int256} vs. \sv{bytes32}
\end{itemize}

The first two cases are treated by initially assigning variables to the most restricted type (\sv{bool}, \sv{address})
and replacing it with the respective \sv{uint} inference if the storage variable ends up being used in integer arithmetic.

The last case is the most challenging as the 3 possible types have many common supported operations, as can be seen in Table \ref{tab:types}.
In addition we can't take advantage of syntactic information such as variable alignment or length extension operations to get type clues.

We handle this case by first assigning an \sv{any32} type to all 32-byte width variables, and replacing that by any other inference based on the type
constraints propagated to them. If no other type constraints are propagated to the storage variable when our analysis reaches its fixpoint,
we replace the \sv{any32} inference with \sv{uint256}.





\section{Evaluation}
\label{sec:evaluation}

The analysis of \ourtool{}, presented as recursive inference rules, is implemented as a set of recursive Datalog
rules on top of the Gigahorse/Elipmoc framework.

We evaluate \ourtool{} over a diverse set of unique smart contracts deployed on the Ethereum mainnet.
To make the evaluation systematic, we take advantage of the \sv{storageLayout} json field output by the Solidity compiler
since version] 0.5.13 \cite{solidity0513}, released in 2019. This compiler output provides the \emph{ground truth}
for our evaluation.


To evaluate our approach we gather all contracts deployed on the Ethereum mainnet up to block 22,800,000
(proposed on the 28th of June 2025) and deduplicate them,
considering two contracts to be duplicates if they contain the same bytecode modulo constant values.
This query returned 903,805 distinct contracts corresponding to 73,729,326 smart contract deployments.%
\footnote{
  Our initial query returned 903,806 distinct contracts corresponding to 78,621,489 smart contract deployments.
  However 4,892,163 of these deployments corresponded to the empty bytecode. This happens for contracts that are
  created and self-destructed within a single transaction. Since these are never stored on the blockchain's state we cannot
  retrieve their bytecode and we have to disregard them.
} 
We filter the dataset, removing 80,580 distinct contracts that implement ``minimal proxy'' patterns.
These contracts are deployed an enormous number of times and add nothing but noise to an evaluation
like ours, since they have no storage variables. 
Following this collection we identified the contracts that have published, verified source code and
use the appropriate compiler version in each case, for ground truth extraction.

This results in our \emph{ground truth} dataset of 377,132 distinct contracts, corresponding to 1,944,788 deployments on the Ethereum mainnet.

We evaluate \ourtool{} against the state-of-the-art VarLifter tool~\cite{VarLifter}.
As VarLifter's storage layout output is not compatible with the \sv{storageLayout} output of the Solidity compiler,
we parse its textual output and produce \sv{solc}-compatible layouts. Additionally, the comparison to the ground truth
for VarLifter is more relaxed than that for \ourtool{}, as VarLifter's output lacks some crucial information:
\begin{itemize}
  \item For storage variables packed into a single slot, no information regarding the 
  offset of each variable is produced.
  \item In the case of struct types that serve as values to mappings, VarLifter does not produce 
  any information about the layout of the struct members.
\end{itemize}

We conducted our experimental evaluation on an idle Ubuntu 24.04 machine with 2 Intel Xeon Gold 6426Y 16 core CPUs and 512G of RAM.
We compile our Datalog analysis using Souffle~\cite{Jordan16,DBLP:conf/cc/ScholzJSW16,souffleInterpreted} version 2.4.1, with 32-bit integer arithmetic and openmp disabled.
An execution cutoff of 300s is used for both tools. \ourtool{} runs 30 single-threaded analysis jobs in parallel, taking advantage
of the native parallelization of the Gigahorse/Elipmoc framework. (This is only a disadvantage for the timings of \ourtool{},
since it may introduce minor contention.) VarLifter, lacking such support, is executed sequentially.

Our evaluation examines analysis performance on the axes of scalability, precision, and completeness (recall). The metrics have standard
definitions, given the ground truth (i.e., the compiler's output): precision is the fraction $\#\mathit{Success}/\#\mathit{Reports}$, while recall
is the fraction $\#\mathit{Success}/\#\mathit{GroundTruth}$. Therefore a missing inference lowers recall, while a wrong inference lowers both precision
and recall.

\subsection{Scalability}

\begin{table}
  \caption{Analysis execution statistics for \ourtool{} on the ground truth dataset.}
  \label{tab:execution:ours}
  \begin{tabular}{|c|c|}
    \hline
            & \ourtool{}  \\
    \hline
    Analysis Terminated & 374,959 (\textbf{99.42\%}) \\
    \hline
    Timeouts            & 2,173   (\textbf{0.58\%})  \\
    \hline
    Errors              & 0 \\
    \hline
    Total               & 377,132 \\
    \hline
    \end{tabular}
\end{table}

Table \ref{tab:execution:ours} shows the execution statistics of \ourtool{} for the full ground truth dataset of 377,132 distinct contracts.
\ourtool{} is able to analyze 99.42\% of contracts in the dataset, under the given 300s execution cutoff.

\begin{table}
  \caption{
    \ourtool{} execution breakdown. Reported decompilation/inline/analysis time excludes the 2,173 contracts that timed out.
  }
  \label{tab:analysis-time}
  \begin{tabular}{|c|c|c|c|}
    \hline
    \ourtool{} analysis stage & Time (secs) &  Timeouts   \\
    \hline
    Decompilation             & 589,757     & 2,146        \\
    \hline
    Inline                    & 581,755     &  0        \\
    \hline
    \ourtool{} analysis       & 186,999     &  27        \\
    \hline
    Total                     & 1,358,511   & 2,173       \\
    \hline
    \end{tabular}
\end{table}

Table \ref{tab:analysis-time} gives more insights into \ourtool{}'s performance.
Based on the time summary for the 374,959 contracts analyzed, \ourtool{} takes an average of 3.62 seconds
per contract.
It is important to note that \ourtool{}'s analysis execution is only accounting for 13.76\% (0.5s on average) of the 
total execution time---the rest is spent on the underlying framework's decompilation and inlining stages.
Additionally, the storage analysis of \ourtool{} introduces very few additional timeouts,
with the vast majority of timeouts coming from the underlying Gigahorse/Elipmoc decompilation stage.

\begin{table}
  \caption{Analysis execution statistics for \ourtool{} and VarLifter on the trimmed dataset.}
  \label{tab:execution:both}
  \begin{tabular}{|c|c|c|}
    \hline
            & \ourtool{} & VarLifter \\
    \hline
    Analysis Terminated & 3748 (\textbf{99.36\%}) & 1906 (\textbf{50.53\%}) \\
    \hline
    Timeouts            & 24   (\textbf{0.64\%})  & 1064 (\textbf{28.20\%}) \\
    \hline
    Errors              & 0                       & 802 (\textbf{21.26\%}) \\
    \hline
    Total               & 3772                    & 3772 \\
    \hline
    \end{tabular}
\end{table}

For comparing against VarLifter we soon realized that, due to scalability limitations, we would not be able to run VarLifter
on our full ground truth dataset. To overcome this we introduce a \emph{trimmed} version of our ground truth dataset containing
1\% of its original contracts, sampled randomly.
Table \ref{tab:execution:both} shows the execution statistics of \ourtool{} and VarLifter for the \emph{trimmed} version of the ground truth
dataset, consisting of 3772 contracts.
\ourtool{} is able to successfully analyze nearly all contracts in this dataset.
On the other hand VarLifter is able to successfully analyze just over half of the contracts.
This is informative, since the VarLifter publication \cite{VarLifter} does not include any statistics on the tool's timeouts and errors.%
\footnote{
  We have also confirmed the high timeout and error rates on the VarLifter publication's own dataset and artifact.
  We have publicly reported this discrepancy along with other issues we identified in the publication's evaluation via github issues
  \url{https://github.com/wsong-nj/VarLifter/issues/1} and \url{https://github.com/wsong-nj/VarLifter/issues/2}.
}

\subsection{Precision}

To measure analysis precision we consider an analysis inference to be successful if it
is able to infer all the variables in a storage slot and their exact types.
It should be noted that achieving a successful inference becomes more difficult as the nestedness and complexity of the defined variables increase.

For example, the nested mapping defined at slot 0x4 in Figure~\ref{fig:storage_layout} requires:

\begin{itemize}
    \item Identifying that it is a 2-nested mapping
    \item Recovering both key-types based on the success criteria
    \item Recovering the value's struct type based on the success criteria
\end{itemize}

\begin{table}[htb]
  \caption{Analysis results for \ourtool{} on the 374,959 deduplicated contracts it analyzed.}
  \label{tab:results:our}
\begin{tabular}{|l|l|}
  \hline
                                 & Result  \\
  \hline
  Ground Truth             & $\leq$ 5,329,677 \\
  \hline
  \ourtool{} Reports             & 5,044,373 \\
  \hline
  \ourtool{} Success       & 4,827,641 \ (Precision \textbf{95.70\%}, Recall $\geq$ \textbf{90.58\%}) \\
  \hline
\end{tabular}
\end{table}

Table \ref{tab:results:our} contains the analysis results of \ourtool{} for the 374,959 distinct contracts
it successfully analyzes. We are focusing on the Precision numbers, i.e., the percentage of
\ourtool{} inferences that also appear in the ground truth.
We can see that \ourtool{} is a very precise analysis, able to infer a variable's exact type 95.7\% of the time.

Next, we consider how \ourtool{} compares against the state-of-the-art VarLifter. Since VarLifter
times out for nearly 50\% of contracts, we perform the precision comparison over the 1896 contracts
that both tools managed to analyze. Table \ref{tab:results:both} shows the results.

\begin{table}[htb]
  \caption{Results for \ourtool{} and VarLifter on the 1896 contracts (subset of the trimmed dataset) analyzed by both tools.}
  \label{tab:results:both}
  \begin{tabular}{|l|l|}
    \hline
                                       & Result  \\
    \hline
    Ground Truth                 & $\leq$ 25963 \\
    \hline
    \ourtool{} Reports                 & 24844 \\
    \hline
    \ourtool{} Success           & 23365 \ (Precision \textbf{97.70\%}, Recall $\geq$ \textbf{93.55\%}) \\
    \hline
    VarLifter Reports           & 17485 \\
    \hline
    VarLifter Success      & 13898 \ (Precision \textbf{83.30\%}, Recall $\geq$ \textbf{55.65\%}) \\
    \hline
  \end{tabular}  
\end{table}

\ourtool{} manages to perform even better for this subset of contracts, successfully identifying the exact types in 97.70\% of reported variables.
On the other hand, VarLifter is significantly less precise, at 83.30\%. Notably, VarLifter's output can
be self-evidently imprecise, reporting colliding types for the same slot.

\subsection{Completeness}
\label{sec:completeness}

We evaluate the completeness of \ourtool{} by examining its ability to recover the ground truth, i.e.,
the \emph{recall} of the analysis: the percentage of variables in the ground truth that \ourtool{} recovers.
Table \ref{tab:results:our} shows the recall of the \ourtool{} to be 90.58\% for the 374,959 contracts of the
ground truth dataset it was able to analyze.

However, \emph{this number is only a lower bound}.

The reason is that real-world smart contracts often need to declare unused variables. These
variables are available to the compiler's ground truth (since the compiler has access to the source
code) but cannot be detected by any bytecode-level analysis. (Inferring these unused variables
is a no-op for all practical purposes.)

The principal case of contracts that declare unused variables is upgradable proxy contracts.
Upgradable proxy contracts need to maintain backwards compatibility of their storage layouts throughout their upgrades.
(Failure to do this can result in storage collisions, a well-recognized problem, also studied in past literature~\cite{NotYourType}.)
The need to maintain compatible storage layouts makes developers continue to declare variables that are no
longer used. Additionally, to avoid storage collisions, developers (and standard upgradability libraries) preemptively declare unused
static arrays in storage, in order to keep a distance between the variables of a Solidity contract and those
of the sub-contracts that inherit from it, so that future versions of the super-contract can add more variables.

One can observe from Table \ref{tab:results:our} that the majority of the incompleteness is reflected in the many fewer
instances of variables inferred by \ourtool{} relative to the ground truth: the ground truth contains
285,304 (5.35\%) more variables, numerically.

We manually inspected 50 randomly-selected instances of such variables missed by \ourtool{},
and all of them turned out to be unused variables.
Our sampling (of 50 out of 5,329,677) has a margin of error of 13.86\% for a confidence level of 95\%.
\footnote{One can verify with a standard \href{https://www.surveyking.com/help/margin-of-error-calculator}{margin-of-error calculator}.}
That is, with 95\% confidence \ourtool{} misses at most 39,543 variables for the contracts in our dataset,
with the rest of the 245,761 reported missing variables being unused ones.
Based on the above, the ground truth includes 5,083,916 variables instead of the 5,329,677 reported by \sv{solc}.
Thus, with 95\% confidence, the real recall of \ourtool{} is \emph{at least} 94.96\%.

Yet another way to appreciate the completeness of \ourtool{} is by comparing the analysis recall to that of VarLifter.
Table \ref{tab:results:both} shows the recall results for both tools, on the contracts in the \emph{trimmed}
dataset that are analyzed by both tools.
\ourtool{} has a significantly higher recall than VarLifter, successfully identifying (at least) 93.55\% of declared contract
variables and their exact types compared to VarLifter's 55.65\%.

The incompleteness of VarLifter is due to its incomplete path extraction algorithm that will
not attempt to visit all code paths. In contrast, the \ourtool{} analysis is recursive to arbitrary depth,
yet fully scalable---e.g., avoiding non-termination issues via the subtle treatment described in Section~\ref{sec:filtering}.

Furthermore, VarLifter's implementation is heavily limited in terms of supported
structures and 
their composability:
\begin{itemize}
  \item Nested mappings can only have a maximum depth of 2.
  \item Nested static arrays can only have at most 2 dimensions.
  \item Only value type and string arrays are supported.
  \item Only structs with members of value types are supported for mapping values. (Whereas storage slots of struct types may also contain strings.)
\end{itemize}

In contrast, the \ourtool{} inference algorithm of Section~\ref{sec:structure-identification} is
capable of detecting arbitrarily-nested storage structures.

\subsection{Practical Value}

To demonstrate the practical value of our storage modeling analysis we show how \ourtool{}
can benefit downstream applications and analyses, deployed on the whole chain:

\begin{enumerate}
  \item \ourtool{} can recover hundreds of thousands of variables missed by the compiler.
  \item \ourtool{} is invaluable for clients analyses, such as the identification of reentrancy guards---a key component of reentrancy analyses.
\end{enumerate}

As context, it is worth noting that \ourtool{} applies to all
deployed contracts on the Ethereum blockchain (and not just the ground
truth dataset of our evaluation, which consisted of contracts with
available source code) and it successfully analyzes \textbf{99.50\%}
of them (899,278 out of 903,805 distinct bytecodes), corresponding to
73,721,169 deployed contracts.

\subsubsection{Incompleteness in the compiler-produced metadata}
\label{sec:compilerIncompleteness}
As discussed in Section~\ref{sec:compiler-patterns}, common low-level storage patterns are not included in
the compiler-produced \sv{storageLayout} json. Therefore, \ourtool{} often retrieves \emph{more} storage
variables than the compiler itself. Of course, the compiler misses these variables because of the use of inline
assembly. However, the inline assembly information is still available to the compiler, and
certainly in much more accessible form than that available to a bytecode-only analyzer.

To quantify the impact of the storage variables missed by the compiler-produced metadata we identified 7 cases of
low-level storage slots used in either EIP/ERC standard proposals or popular libraries. Each of these code patterns
contains variables that the compiler misses, yet \ourtool{} (overwhelmingly) detects.\footnote{Although there is no
ground truth for variables that the compiler misses, the performance of \ourtool{} for them is expected to
be similar to that for variables that the compiler does know about. Patterns such as those of Figure~\ref{fig:lowLevel}
are not a problem for a bytecode-level analyzer, like \ourtool{}. In manual sampling of contracts with inline assembly, we have found no
evidence of different \ourtool{} precision or recall for these variables.}
We identify these patterns in all deployed contracts and present statistics on their frequency in
Table \ref{tab:analytics-deployments}.

Thus, \ourtool{} is able to recover hundreds of thousands of instances of low-level storage variables deployed in 
the Ethereum mainnet.

\begin{table}[]
\centering
\caption{On-chain usage of various standard patterns using low-level storage variables.}
\begin{tabular}{|l|r|r|}
\hline
\textbf{Pattern} & \textbf{\# Distinct} & \textbf{\# Deployments} \\
\hline
Diamond Proxy & 506 & 1,147 \\
1967 Proxy & 15,972 & 257,180 \\
1967 Beacon & 208 & 93,820 \\
1822 Proxy & 359 & 23,331 \\
OZ Proxy & 121 & 2,836 \\
OZ Initializable & 8,213 & 10,928 \\
OZ ReentrancyGuardUpgr & 2,409 & 3,392 \\
\hline
\end{tabular}
\label{tab:analytics-deployments}
\end{table}

\subsubsection{Client: Reentrancy Guards}

As an example of a client analysis that clearly relies on \ourtool{}, we consider a detector of reentrancy guards.
The typical reentrancy guard pattern, shown in Figure~\ref{fig:guard}, treats a special storage variable
as a mutex, using it to prevent reentry to the contract's other protected functions. 

\begin{figure}
\begin{solidity}
modifier nonReentrant() {
  require(_status == NOT_ENTERED);
  _status = ENTERED;
  _; // function code goes here
  _status = NOT_ENTERED;
}
\end{solidity}
\caption{Standard reentrancy guard pattern}
\label{fig:guard}
\end{figure}

The identification of reentrancy guards is instrumental for increasing the precision of static reentrancy analyses by
considering calls between the two guard-setting statements as reentrancy-safe.

We implement a detector for reentrancy guards in 70 lines of Datalog code, combining the storage analysis of \ourtool{}
with value-flow, control-flow, and data-flow predicates and components provided by the Gigahorse framework~\cite{gigahorse}.

The checker identifies reentrancy guards in 187,633 distinct contract bytecodes, corresponding to 811,639 contracts deployed
on the Ethereum mainnet.

\section{Related Work}

Reasoning about the usage of the EVM's storage has been instrumental for analysis tools and decompilers.
However, no past tools (other than VarLifter---extensively compared earlier) attempt
to statically fully recover storage structures as-if in the source program. For instance, past analyses may have
inferred ``this is an access to the \sv{balances} mapping'' but not the width of an entry, the full
nested structure of the mapping, or the type of elements. Such work, discussed next, can potentially
benefit from our techniques.

Most early frameworks \cite{Albert2018,Tsankov:2018:SPS:3243734.3243780} would only precisely
reason about storage loading / storing statements with indexes resolved to constant values,
sacrificing precision or completeness in low-level code treating dynamic data structures.
Madmax \cite{MadMax} was the first work to propose an analysis that inferred high-level
structures (arrays and mappings) from EVM bytecode. This analysis enabled MadMax
to detect storage-related vulnerabilities focusing on griefing and DoS. Ethainter \cite{ethainter}
also made use of the storage modeling introduced in \cite{MadMax} to detect guarding patterns and track
the propagation of taint through storage. An implicit modeling of storage was also achieved (as \sv{keccak256} expressions with free variables) in \cite{symvalic}.

In addition, as \ourtool{} is publicly available in an open-source repository, the results of previous snapshots of our codebase
have been incorporated into independently published research tools.
One such example is BlockWatchdog~\cite{BlockWatchdog}, which leverages \ourtool{}'s storage modeling to identify storage variables that hold addresses of external contracts.

The recent CRUSH tool ~\cite{NotYourType} implements a storage collision vulnerability analysis for 
upgradable proxy contracts. Part of this work involves modeling storage, including a modeling of mappings, arrays,
and byte-ranges of constant-offset storage slots to discriminate between storage variables packed into a single slot.
Unfortunately, the work lacks support for arbitrarily-nested data structures.
These same techniques have been applied (for the same security application) to the Proxion tool \cite{proxion}.
Another recent tool~\cite{AlbertStorage} analyzes storage access patterns to precisely compute the gas
bounds of contracts via a Max-SMT based approach.

Other work has focused on analyzing the usage patterns of the EVM's various ``memory'' stores.
\cite{ethereum-memory-oopsla} proposes techniques to infer high-level facts from EVM bytecode.
These inferences include high-level uses of operations reading from memory (hashing operations, external calls)
memory arrays and their uses. \ourtool{} relies on these inferences for the modeling of the storage index values,
and the propagation of type constraints through memory.
The Certora prover employs a memory splitting transformation~\cite{CertoraMemory} after modeling
the allocations of high-level arrays and structs of EVM contracts and the aliasing between different allocations.
This transformation allows the tool to consider disjoint memory locations separately, speeding up SMT queries by up to 120x.
In other work~\cite{AlbertMemory} the uses of memory are analyzed to identify optimization opportunities.

Several other end-to-end applications rely on storage modeling. Storage modeling is also used in blockchain explorers and can be done dynamically.
The now-discontinued storage explorer tool, \href{https://evm.storage}{evm.storage},
used such a dynamic analysis, by examining the use of hash pre-images derived from past executions of the contract.
Analysis of confused deputy attack contracts has employed both static and dynamic storage modeling techniques~\cite{ConfusumContractum}. 
A smart contract policy enforcer, \textsc{EVM-Shield}, utilizes storage modeling to pinpoint functions that perform state updates and
adds pre- and post-conditions within the smart contract itself to prevent malicious transactions on-chain~\cite{EVM-Shield}.

Related to our work, past tools~\cite{SigRec,DeepInfer} have been proposed to infer the ABI interfaces of unknown contracts by inferring
the structures and types of public function arguments. Such tools can benefit from our work by taking our inferences into account in
their type recovery efforts.
As an example, SigRec~\cite{SigRec}, lacking a model of the EVM's storage, considers any variable read from or written to storage to be
of type \sv{uint256}.

Outside the domain of smart contracts, a number of techniques on variable recognition and type inference of binaries are relevant to our work ~\cite{lee2011tie,10.1145/2896499,9153442,PrologRecover,FindingDwarf,10.1145/2499370.2462165,TypeBased,DIVINE,10.1134/S0361768809020066}.
Static-analysis-based approaches \cite{TypeBased,DIVINE,AggregateStructure} have historically seen widespread adoption in this setting.
Recent learning-based tools \cite{TypeFSL,TypeMiner,StateFormer} have also been successful in recovering type information from binaries.
More closely related to our approach, OOAnalyzer \cite{PrologRecover} uses Prolog to infer C++ classes from binaries.

\section{Conclusion}

We presented \ourtool{}, a static-analysis-based lifter for storage variables from the binaries of Ethereum smart contracts.
\ourtool{}, powered by an analysis of low-level storage indexes, is able to resolve arbitrarily-nested
high-level data structures from the low-level bytecode.
Compared against the state-of-the-art in a diverse dataset of real-world contracts, \ourtool{} manages to excel in all
evaluation dimensions: scalability, precision, and completeness.
Analyzing all deployed contracts, \ourtool{} identifies variables missed
by the Solidity compiler in hundreds of thousands of deployed contracts.

\begin{acks}
We gratefully acknowledge funding by ERC Advanced Grant \textsc{PINDESYM} (101095951).
\end{acks}

\section*{Data-Availability Statement}
The implementation of \ourtool{} is available in the \url{https://github.com/nevillegrech/gigahorse-toolchain} open source repository.
The paper's artifact~\cite{StorageArtifact} is available on Zenodo.

\bibliographystyle{ACM-Reference-Format}
\bibliography{bibliography,tools,references}

\newpage
\appendix

\section{(Packed) Variable Partitioning Analysis}
\label{sec:packed}

To identify instances of multiple high-level variables taking up part of and sharing the same EVM storage word, we need to model
their uses in high-level operations as well as their definitions.

Figure ~\ref{rels2} contains additional input relations needed to identify instances of these packed variables.
Predicates handling masking and shifting are convenience wrappers for low-level arithmetic and bitwise operations of variables and constants:
\sv{LowBytesMask} and \sv{HighBytesMask} correspond to operation patterns that use \sv{AND} instructions; \sv{LShift} to patterns with \sv{MUL} and \sv{SHL} (a left-shift); \sv{RShift} to patterns with \sv{DIV} and \sv{SHR} (right-shift).

\begin{figure}[]
  \begin{tabular}{|l|}
    \hline
    \Flows{f}{t} $|$ \myvar{f} $\in$ $V$, \myvar{t} $\in$ $V$ \\
    Limited data flows analysis: \myvar{f} flows to \myvar{t} through low-level \\shifting and masking operations. \\
    \hline
    \statement{LowBytesMask}{u, w}{r}  $|$ \myvar{r} $\in$ $V$, \myvar{u} $\in$ $V$, \myvar{w} $\in$ \type{Int} \\
    Masking operation (\myvar{w} bytes of mask width) used in  casting \myvar{u} \\ to value types  \sv{uintX}, \sv{intX}, \sv{address}, \sv{bool} \\
    \hline
    \statement{HighBytesMask}{u, w}{r} $|$ \myvar{r} $\in$ $V$, \myvar{u} $\in$ $V$, \myvar{w} $\in$ \type{Int} \\
    Masking operation (\myvar{w} bytes of mask width) used in casting \myvar{u} \\to the \sv{bytes}$X$ value types \\
    \hline
    \statement{LShift/RShift}{u, n}{r} $|$ \myvar{r} $\in$ $V$, \myvar{u} $\in$ $V$, \myvar{n} $\in$ \type{Int} \\
    Variable \myvar{u} is shifted to the left/right by \myvar{n} bytes. \\
    \hline
    \statement{BooleanCast}{u}{r} $|$ \myvar{r} $\in$ $V$, \myvar{u} $\in$ $V$ \\
    Low-level cast to boolean using two consecutive \sv{ISZERO} \\operations. \\
    \hline
    \pred{HighLevelOpUse}{\myvar{v}} $|$ \myvar{v} $\in$ $V$ \\
    Variable \myvar{v} is used in high-level operation \\(e.g., non-storage-address computation, calls). \\
    \hline
  \end{tabular}
  \caption{Additional input relation definitions}
  \label{rels2}
  \end{figure}

The following relations capture the inferences of the analysis.

\begin{center}
  \begin{tabular}{|l|}
  \hline
  \StorageVarBytesUse{\myvar{sv}}{\mystmtid{I}}{\myvar{v}}{\myvar{l}}{\myvar{h}} $|$ \mystmtid{I} $\in$ \type{S}, \myvar{v} $\in$ \type{V}, $\mathsf{sv}$ $\in$ \Variable{}, \myvar{l} $\in$ \type{Int}, \myvar{h} $\in$ \type{Int} \\
  \textbf{Partial Read}: \\Variable \myvar{v} holding byte range [\myvar{l} : \myvar{h}] of storage variable $\mathsf{sv}$, \\loaded via \mystmtid{I},
  is used in high-level operation or is not \\cast further. \\
  \hline
  \StorageVarBytesDef{\myvar{sv}}{\mystmtid{I_S}}{\mystmtid{I_L}}{\myvar{v}}{\myvar{l}}{\myvar{h}}  $|$ \mystmtid{I_S} $\in$ \type{S}, \mystmtid{I_L} $\in$ \type{S}, $\mathsf{sv}$ $\in$ \Variable{},\\ \myvar{l} $\in$ \type{Int}, \myvar{h} $\in$ \type{Int}, \myvar{v} $\in$ \type{V} \\
  \textbf{Partial Write}:\\ Variable \myvar{v} is written to byte range [\myvar{l} : \myvar{h}] of storage variable $\mathsf{sv}$ \\in statement \mystmtid{I_S}.
  All other contents of $\mathsf{sv}$ are retained as loaded \\in statement $I_L$. \\
  \hline
  \AnyStorageVarBytes{\mystmtid{I}}{$\mathsf{sv}$}{\myvar{l}}{\myvar{h}} $|$ \mystmtid{I} $\in$ \type{S}, $\mathsf{sv}$ $\in$ \Variable{}, \myvar{l} $\in$ \type{Int}, \myvar{h} $\in$ \type{Int} \\
  Aggregation of the two previous relations. \\
  \hline
  \mergefail{\mathsf{sv}} $|$ $\mathsf{sv}$ $\in$ \Variable{} \\
  \textbf{Partitioning Analysis Failure}: \\Packed variable analysis failed to partition $\mathsf{sv}$ into multiple\\ \PackedVariable{} instances. \\
  \hline
  \VarHoldsBytes{\myvar{v}}{\mystmtid{I}}{\myvar{sv}}{\myvar{l'}}{\myvar{h'}}{\myvar{l}}{\myvar{h}} $|$ \mystmtid{I} $\in$ \type{S}, $\mathsf{sv}$ $\in$ \Variable{},\\ \myvar{l'} $\in$ \type{Int}, \myvar{h'} $\in$ \type{Int}, \myvar{v} $\in$ \type{V}, \myvar{l} $\in$ \type{Int}, \myvar{h} $\in$ \type{Int} \\  
  \textbf{Intermediate Partial Read}: Bytes [\myvar{l'} : \myvar{h'}] of variable \myvar{v} \\ hold bytes [\myvar{l} : \myvar{h}] of storage variable $\mathsf{sv}$, loaded via \mystmtid{I}.\\
  \hline
  \end{tabular}
\end{center}

At a first approximation, the analysis merely
tracks constant-offset additions and constant-mask boolean operations that the compiler outputs. 
The rules of this appendix are to some extent just tedious ``work''.
However, we explicitly list the rules/patterns recognized for technical concreteness and completeness,
especially for detail-oriented readers who may question what pattern recognition
can reliably yield the results reported in the experiments in Section \ref{sec:evaluation}.

\subsection{Uses of Packed Variables}

Figure~\ref{fig:VarHoldsBytes} shows how the contents of a storage variable are tracked through
sequences of shifting and casting operations. All rules are recursive, with the base case
of the recursion being that an \textit{intermediate} partial read fact ``\VarHoldsBytes{\mathsf{v}}{\mystmtid{I_L}}{\mathsf{sv}}{0}{31}{0}{31}''
is produced for each \sv{LOAD} statement loading an index corresponding to storage variable $\mathsf{sv}$.

\begin{figure}
\[
\textsc{(Base)}
\inferrule{
  \MapToStorVar{\mathit{I_L}}{\mathsf{sv}} \\
  \mathit{I_L}: [\mathsf{v} := \mathsf{LOAD}(\_)]
}{
  \VarHoldsBytes{\mathsf{v}}{\mathit{I_L}}{\mathsf{sv}}{\mathsf{0}}{\mathsf{31}}{\mathsf{0}}{\mathsf{31}}
}
\]

\[
\textsc{(RShift)}
\inferrule{
  \VarHoldsBytes{\mathsf{pv}}{\mathit{I_L}}{\mathsf{sv}}{\mathsf{s}}{\mathsf{31}}{\mathsf{l}}{\mathsf{h}} \\
  \mathsf{v} := \mathsf{RShift}(\mathsf{pv}, \mathsf{n})
}{
  \VarHoldsBytes{\mathsf{v}}{\mathit{I_L}}{\mathsf{sv}}{\max(\mathsf{s} - \mathsf{n}, 0)}{\mathsf{31}}{\mathsf{l} + \mathsf{n} - \mathsf{s}}{\mathsf{h}}
}
\]

\[
\textsc{(LShift1)}
\inferrule{
  \VarHoldsBytes{\mathsf{pv}}{\mathit{I_L}}{\mathsf{sv}}{\mathsf{s}}{\mathsf{31}}{\mathsf{l}}{\mathsf{h}} \\
  \mathsf{v} := \mathsf{LShift}(\mathsf{pv}, \mathsf{n}) \\
  \mathsf{w}  := 1 + \mathsf{h} - \mathsf{l} \\
  \mathsf{s} + \mathsf{n} + \mathsf{w} \leq 32
}{
  \VarHoldsBytes{\mathsf{v}}{\mathit{I_L}}{\mathsf{sv}}{\mathsf{s} + \mathsf{n}}{\mathsf{31}}{\mathsf{l}}{\mathsf{h}}
}
\]

\[
\textsc{(LShift2)}
\inferrule{
  \VarHoldsBytes{\mathsf{pv}}{\mathit{I_L}}{\mathsf{sv}}{\mathsf{s}}{\mathsf{31}}{\mathsf{l}}{\mathsf{h}} \\
  \mathsf{v} := \mathsf{LShift}(\mathsf{pv}, \mathsf{n}) \\
  \mathsf{w}  := 1 + \mathsf{h} - \mathsf{l} \\
  \mathsf{s} + \mathsf{n} + \mathsf{w} > 32
}{
  \VarHoldsBytes{\mathsf{v}}{\mathit{I_L}}{\mathsf{sv}}{\mathsf{s} + \mathsf{n}}{\mathsf{31}}{\mathsf{l}}{\mathsf{h} - (\mathsf{s} + \mathsf{n} + \mathsf{w} - 32)}
}
\]

\[
\textsc{(LBMask)}
\inferrule{
  \VarHoldsBytes{\mathsf{pv}}{\mathit{I_L}}{\mathsf{sv}}{\mathsf{s}}{\mathsf{31}}{\mathsf{l}}{\mathsf{h}} \\
  \mathsf{v} := \mathsf{LowBytesMask}(\mathsf{pv}, \mathsf{m}) \\
  \mathsf{s} < \mathsf{m}
}{
  \VarHoldsBytes{\mathsf{v}}{\mathit{I_L}}{\mathsf{sv}}{\mathsf{s}}{\mathsf{31}}{\mathsf{l}}{\min(\mathsf{s} + \mathsf{h}, \mathsf{s} + \mathsf{l} + \mathsf{m} - 1) - \mathsf{s}}
}
\]

\[
\textsc{(HBMask)}
\inferrule{
  \VarHoldsBytes{\mathsf{pv}}{\mathit{I_L}}{\mathsf{sv}}{\mathsf{s}}{\mathsf{31}}{\mathsf{l}}{\mathsf{h}} \quad
  \mathsf{v} := \mathsf{HighBytesMask}(\mathsf{pv}, \mathsf{m}) \\
}{
  \VarHoldsBytesSplitA{\mathsf{v}}{\mathit{I_L}}{\mathsf{sv}}{\max(32-\mathsf{m},\mathsf{s})}{\mathsf{31}}\\
  \VarHoldsBytesSplitB{\mathsf{sv}}{\max(\mathsf{s} + \mathsf{l}, \mathsf{s} + \mathsf{h} - \mathsf{m} + 1) - \mathsf{s}}{\mathsf{h}}
}
\]

\[
\textsc{(BoolCast)}
\inferrule{
  \VarHoldsBytes{\mathsf{pv}}{\mathit{I_L}}{\mathsf{sv}}{\mathsf{0}}{\mathsf{31}}{\mathsf{b}}{\mathsf{b}} \\
  \mathsf{v} := \mathsf{BooleanCast}(\mathsf{pv})
}{
  \VarHoldsBytes{\mathsf{v}}{\mathit{I_L}}{\mathsf{sv}}{\mathsf{0}}{\mathsf{31}}{\mathsf{b}}{\mathsf{b}}
}
\]
\caption{Tracking storage variables through casts and shifts}
\label{fig:VarHoldsBytes}
\end{figure}

When this computation reaches fixpoint, intermediate inferences are promoted to full partial read inferences based on the criteria shown in Figure \ref{fig:partWrite}.
Rule \textsc{Use1} will infer a partial read when the variable holding an intermediate inference is not cast or shifted further,
while also ensuring it does not flow to a STORE statement through low-level casting and shifting operations, eliminating LOAD statements
used in partial \textit{write} patterns. The \textsc{Use2} rule validates intermediate inferences held by variables used in high-level operations,
while the \textsc{Use3} rule increases completeness by validating intermediate inferences of \Variable{} subregions that have been used in other partial read operations.

\begin{figure}
\[
\textsc{(Use1)}
\inferrule{
  \VarHoldsBytes{\myvar{v}}{\mathit{I_L}}{\mathsf{sv}}{\_}{\_}{\mathsf{l}}{\mathsf{h}} \\
  \nexists I_S: \big(\MapToStorVar{I_S}{\mathsf{sv}} \quad I_S: \mathsf{STORE}(\mathsf{u}) := \_ \quad \Flows{\mathsf{v}}{\mathsf{u}} \big) \\
  \neg\, \_ := \mathsf{LShift}(\myvar{v},\_) \quad
  \neg\, \_ := \mathsf{RShift}(\myvar{v},\_) \\
  \neg\, \_ := \mathsf{LowBytesMask}(\myvar{v},\_) \quad
  \neg\, \_ := \mathsf{HighBytesMask}(\myvar{v},\_) \\
}{
  \StorageVarBytesUse{\mathsf{sv}}{\mathit{I_L}}{\mathsf{v}}{\mathsf{l}}{\mathsf{h}}
}
\]

\[
\textsc{(Use2)}
\quad
\inferrule{
  \VarHoldsBytes{\mathsf{v}}{\mathit{I_L}}{\mathsf{sv}}{\_}{\_}{\mathsf{l}}{\mathsf{h}} \\
  \mathsf{HighLevelOpUse}(\mathsf{v}) \\
}{
  \StorageVarBytesUse{\mathsf{sv}}{\mathit{I_L}}{\mathsf{v}}{\mathsf{l}}{\mathsf{h}}
}
\]

\[
\textsc{(Use3)}
\quad
\inferrule{
  \VarHoldsBytes{\mathsf{v}}{\mathit{I_L}}{\mathsf{sv}}{\_}{\_}{\mathsf{l}}{\mathsf{h}} \\
  \StorageVarBytesUse{\mathsf{sv}}{\_}{\_}{\mathsf{l}}{\mathsf{h}} \\
}{
  \StorageVarBytesUse{\mathsf{sv}}{\mathit{I_L}}{\mathsf{v}}{\mathsf{l}}{\mathsf{h}}
}
\]
\caption{Inferring partial storage read inferences.}
\label{fig:partWrite}
\end{figure}

\subsection{Stores of Packed Variables}
The above rules inferred use of packed variables from \emph{uses} of the variables, i.e., from \sv{LOAD} statements and subsequent patterns. Similar logic applies to the \emph{definitions} of the variables, i.e., code
that writes to a storage word via a \sv{STORE} statement.

The \StorageVarBytesDef{\myvar{sv}}{I_S}{I_L}{\myvar{v}}{\myvar{l}}{\myvar{h}} ``partial write'' (by analogy to the earlier
``partial read'') is the result of the analysis tracking the stores of packed variables.
We do not show the tedious rules explicitly, but briefly they track variables
through the following low-level patterns:
\begin{enumerate}
  \item The contents of a storage variable $\mathsf{sv}$ are loaded by the $I_L$ statement.
  \item The $[\myvar{l}:\myvar{h}]$ range of bytes is masked off, disregarding their contents.
  \item The new value, held in variable $\mathsf{v}$, is shifted into the correct byte offset, if it happens to not  already be at the correct offset.
  \item The shifted variable and the contents of the other variables occupying the same slot are combined using a bitwise \sv{OR} operation.
  \item The result of the previous step is stored to $\mathsf{sv}$ in statement $I_S$.
\end{enumerate}

It should be noted that, in optimized code, multiple nearby writes will be grouped, resulting in multiple
partial write inferences for the same $(\mathsf{sv}, I_S, I_L)$ but different byte ranges, corresponding to different packed variables.

\subsection{Inference aggregation}

After computing the reads and writes of possibly packed storage \Variable{} instances,
we aggregate their results to ensure they do not contain conflicting inferences.

\begin{figure}
\[
\textsc{(Conflict1)}
\inferrule{
    \AnyStorageVarBytes{\mathit{I}}{\mathsf{sv}}{\mathsf{l}}{\mathsf{h}} \quad
    \AnyStorageVarBytes{\mathit{I'}}{\mathsf{sv}}{\mathsf{l'}}{\mathsf{h'}} \\
    (\mathsf{l}, \mathsf{h}) \neq (\mathsf{l'}, \mathsf{h'}) \quad
    (\mathsf{l}, \mathsf{h}) \cap (\mathsf{l'}, \mathsf{h'}) \neq \emptyset
}{
  \mergefail{\mathsf{sv}}
}
\]

\[
\textsc{(Conflict2)}
\inferrule{
    \AnyStorageVarBytes{\mathit{I}}{\mathsf{sv}}{\mathsf{l}}{\mathsf{h}} \quad
    \AnyStorageVarBytes{\mathit{I'}}{\mathsf{sv}}{\mathsf{l'}}{\mathsf{h}} \quad
    \mathsf{l} \neq \mathsf{l'}
}{
  \mergefail{\mathsf{sv}}
}
\]

\[
\textsc{(Missing)}
\quad
\inferrule{
  \AnyStorageVar{\mathit{I}}{\mathsf{sv}} \\  
  \neg\, \StorageVarBytesUse{\mathsf{sv}}{\mathit{I}}{\_}{\_}{\_} \\
  \neg\, \StorageVarBytesDef{\mathsf{sv}}{\mathit{I}}{\_}{\_}{\_}{\_} \\
  \neg\, \StorageVarBytesDef{\mathsf{sv}}{\_}{\mathit{I}}{\_}{\_}{\_} \\
}{
  \mergefail{\mathsf{sv}}
}
\]

\caption{Logic identifying conflicting inferences of \Variable{} sharing the same slot.
If the failed variable partitioning inference ($\mergefail{}$) is not produced for a given storage variable $\mathsf{sv}$ it is considered successfully merged
and all inferred partial reads and writes are matched with their corresponding \PackedVariable{} instances.}
\label{fig:aggr}
\end{figure}

Figure~\ref{fig:aggr} captures the cases when this inference \emph{fails}, i.e., 
when the analysis infers conflicting offsets for the same variable, or does not
manage to infer any offsets for a variable. The partitioning analysis failure ($\mergefail{}$) predicate
of Figure~\ref{fig:aggr} is used negatively: if it does not apply, the packed variable
analysis has been successful (for the specific variable being considered)---there is
an inference of an offset inside a storage word and the offset is unique.
{}

\end{document}